\documentstyle[epsfig]{aipproc}

\begin{document}
\title{Interpreting CMB Anisotropy Observations:  Trying to Tell 
the Truth with Statistics}

\author{Eric Gawiser}
\address{Center for Astrophysics and Space Sciences, University 
of California, San Diego\\La Jolla, CA  92093}

\maketitle

\begin{abstract}
A conflict has been reported between the baryon density 
inferred from deuterium observations and that found from recent 
CMB observations by BOOMERanG and MAXIMA.  
Despite the flurry of papers that attempt to resolve this 
conflict by adding new physics to the 
early universe,  we will show that it can instead be resolved via a 
more careful usage of statistics.  
Indeed, the Bayesian analyses that produce this conflict 
are by their nature poorly suited for drawing this type of 
conclusion.  A properly defined frequentist analysis can address this 
question directly and appears not to find a conflict.  
Finally, a conservative accounting of 
systematic uncertainties in measuring the 
deuterium abundance could reduce what is nominally a $3 \sigma$ conflict 
to $1 \sigma$.  

\end{abstract}

\section*{Introduction}

The recent BOOMERanG \cite{debernardisetal00,langeetal01} 
and MAXIMA-1 \cite{hananyetal00,balbietal00} observations of Cosmic Microwave 
Background (CMB) anisotropy provide the first high-quality, 
high-resolution observations to cover the angular scales over which 
the first two acoustic peaks are expected in the angular 
power spectrum.  If used by themselves, these data are sufficient 
to determine the location and rough amplitude of the first acoustic peak, 
providing evidence that the universe is near critical density.  A simultaneous 
fit to numerous cosmological parameters is impossible, however, because 
of strong degeneracies amongst those parameters in determining the 
shape of the CMB angular power spectrum at the precision of the 
observations.  As a result, a Bayesian framework has been used in which 
numerous other cosmological observations are used as priors.  These 
priors come from large-scale structure, 
Type Ia Supernovae, direct determinations of Hubble's constant, and 
the baryon density inferred from combining observations of the 
deuterium-to-hydrogen abundance ratio 
with the standard predictions of Big Bang Nucleosynthesis (BBN).

A funny thing happened on the way to precision cosmology.  
While the first acoustic peak is clearly defined by CMB anisotropy data, 
thereby providing evidence for a flat universe, the second acoustic peak 
is either at surprisingly low amplitude or missing entirely.  Within 
the parameter space of the standard adiabatic CDM paradigm, this can be 
produced either by a red-tilted ($n<1$) primordial power spectrum or 
by a baryon density higher than that inferred from deuterium observations 
plus BBN ($\Omega_b h^2 = 0.021\pm 0.002$) \cite{omearaetal01}.  In order 
to test the latter idea, \cite{jaffeetal00} performed 
a Bayesian analysis on the combined BOOMERanG,MAXIMA-1, and 
COBE-DMR \cite{bennettetal96} data without using 
the BBN baryon density as a prior, and they found
 $\Omega_b h^2 = 0.032\pm 0.004$.  Hence there appears to be a conflict between
the CMB and BBN values for the baryon density at roughly the $3\sigma$ level.

One way to resolve this conflict is to postulate additional physics 
in the early universe that alters Big Bang Nucleosynthesis such that 
the observed deuterium abundance is consistent with the higher value of 
$\Omega_b h^2$ preferred by the CMB.  This has been attempted using 
degenerate neutrinos due to a large lepton asymmetry 
\cite{lesgourguesp00,espositoetal01}, a decaying neutrino that 
likewise produces extra entropy during BBN \cite{hansenv00,kaplinghatt01},  
or inhomogeneous BBN \cite{kurkisuonios00}.
Even if the precise priors on the nature of nonstandard BBN are allowed 
to vary, a robust need for new physics is claimed \cite{knelleretal01}.  
Another approach adds new physics to the earlier inflationary epoch in the 
form of an unexplained bump in the primordial power spectrum of 
density perturbations \cite{griffithssz00}.  

Injecting new physics into the first few minutes of the 
universe is a serious step and needs to be motivated by a strong observational 
signal.  While the claimed $3 \sigma$ conflict between CMB and BBN baryon 
densities seems to have been interpreted by many authors as a sufficient
signal, 
a close 
examination of the statistics involved reveals that this conflict has 
been exaggerated and may not exist at all.  
 

\section*{Bayesian Analyses}

A Bayesian analysis seeks to answer the question, ``Given what I knew 
before plus the data I have just obtained, what do I now think the 
truth is?''  What was known before is incorporated in the form of a 
prior probability function.  Basic probability theory gives us the starting 
point, 
\begin{equation}
p(model,data) = p(data|model) p(model) = p(model|data) p(data) 
\; , \; \; \; 
\end{equation}
which is equivalent to the statement that the probability of two things 
being true is equal to the probability that the first one is true  
given that the second one is true times the probability that the second 
one is indeed true.  Bayes' theorem involves specializing this statement 
to the case of a set of models and the observed data and dividing 
by $p(data)$ to get
\begin{equation}
p(model|data) \propto p(data|model) p(model) \; . \; \; 
\end{equation}
The probability that the data would be observed given a particular model
is often easy to calculate and is referred to as the likelihood 
function.  This means that as long as we know the 
prior probability of various models being true, $p(model)$, and can 
calculate the likelihood function, we can determine the posterior likelihood 
that each model is correct.  We can either think of deuterium observations 
as part of the current data and do a joint likelihood analysis or we can 
account for the results of the deuterium observations when we choose a prior 
probability function for $\Omega_b h^2$.  To ignore the deuterium observations 
entirely would imply that we do not consider them trustworthy.  

Since the various models considered by \cite{jaffeetal00} all lie 
within the adiabatic CDM parameter space, the prior $p(model)$ can 
be expressed as the product of the 
prior probability functions of various independent parameters, 
including the baryon density.  When a uniform prior 
$0.0031\leq \Omega_b h^2 \leq 0.2$ is used, these authors find a 
 posterior likelihood described by $ \Omega_b h^2 = 0.032 \pm 0.004$.  
If a prior consistent with BBN plus deuterium observations,
$ \Omega_b h^2 = 0.019 \pm 0.002$ \cite{burlest98}, 
is used, they find a posterior 
likelihood described by  $ \Omega_b h^2 = 0.021 \pm 0.003$.  

Although these authors conclude that the BBN plus deuterium value of 
the baryon density is ``disfavored by the data,'' this is 
not the right interpretation to ascribe to the results of their 
Bayesian analysis.  If they assume that the BBN plus deuterium 
value of the 
baryon density is correct by including it as a prior, they produce a 
posterior likelihood in good agreement, showing that while the CMB data 
may favor a higher value of $\Omega_b h^2$, BBN is a much stronger 
constraint.  Indeed, they note that this prior is strong enough to 
alter the results on other parameters, for instance yielding a 
scalar spectral index of the primordial power spectrum of $n_s = 0.89 
\pm 0.06$ rather than the $n_s=1.03\pm0.08$ produced by the uniform 
prior on the baryon density.  Starting and ending with a baryon density 
consistent with BBN plus deuterium is a self-consistent result.  

When they instead use a uniform prior on the baryon density and ignore 
the implications of deuterium observations, these authors are starting 
from the assumption that the BBN value of the baryon density is not 
worthy of consideration.  Producing a posterior likelihood for $\Omega_b 
h^2$ that is different from the BBN value is again self-consistent; 
in this case we start and end with the idea that $\Omega_b h^2$ 
may well be greater than 0.019.  In the case of the uniform prior, the 
CMB data do show the ability to narrow a broad prior into a localized 
posterior.  The correct conclusion to draw from this exercise is that it is 
quite important to decide a priori whether we believe the deuterium 
observations and what they imply for the baryon density, because it 
makes a significant difference in our posterior estimation of the truth.  

One complication of Bayesian statistics is that if we do not 
know the correct priors we should vary our priors over the range of 
reasonable functions.  If indeed both the BBN and the uniform prior on 
the baryon density are reasonable, then the correct conclusion about the 
posterior likelihood is that $\Omega_b h^2$ could be anywhere from 0.02 to 
0.03.  Because it requires a specific choice of prior assumptions and produces 
only relative likelihoods at the end, the simple Bayesian analysis is not 
well-suited to answering the question, ``Are the CMB and BBN values for 
the baryon density in conflict?''  This question could be pursued in a 
Bayesian format using prior assumptions on how likely such a conflict is, 
but this is far beyond the scope of the analyses that have been done.

\section*{Frequentist Analyses}
 
A frequentist analysis is a bit simpler to describe; each model is 
viewed as a separate hypothesis to be tested against the observations.  
One looks only at the likelihood function i.e. $p(data|model)$.  Typically 
a misfit statistic such as $\chi^2$ is used, and any models for which the 
chance of getting a better agreement with the data is greater than 
or equal to e.g. 95\% are considered to be ruled out at the e.g. 
95\% confidence level.  This is more akin to the basic scientific method 
taught to children; a frequentist analysis seeks to answer the question, 
``Which of these models are reasonably likely to produce the observed data?''  
While a best-fit model can still be found, one concentrates on discarding 
those models that are ruled out beyond some confidence level.  Further 
discrimination requires better data.  This frequentist approach has the 
added benefit of being able to rule out an entire parameter space if none 
of its models are a reasonably good fit to the data; in this case the 
hypothesis that the true model lies somewhere in this parameter space has 
been rejected.  The Bayesian approach can be modified to compare one 
parameter space to another but it always makes a conclusion based on 
relative likelihood.  

In the case of the recent CMB data, a frequentist goodness-of-fit analysis 
was performed by the MAXIMA team \cite{balbietal00}.  They find that 
the $\Lambda$CDM model with $\Omega_b h^2 = 0.021$ 
has $\chi^2 = 10/10$ when compared with the MAXIMA-1  
data alone and $\chi^2 = 40/40$ compared with MAXIMA-1 plus COBE-DMR, but their best-fit model with $\Omega_b h^2 = 0.025$ 
has $\chi^2 = 8/10$ and $\chi^2=38/40$ respectively.  
This is consistent with the Bayesian result that 
a high baryon density has greater likelihood, but now we have a chance 
to assess the absolute goodness-of-fit of these models.  The ``best-fit'' 
model is a slightly better fit than one expects but this is likely 
explained by having varied seven parameters to find it.  Indeed, we should 
subtract up to seven degrees of freedom from the above results if these 
seven parameters successfully span the space of possible functions of 
seventh order.  A simple way to state this effect is 
that even if $\Lambda$CDM is the true model we expect that by varying 
$n$ of its parameters freely we will be able to drop the $\chi^2$ by 
a value of $n$; these parameter variations are fitting the 
observational errors in the data rather than telling us more about 
the truth.  

Of course, we would like to have a similar frequentist analysis of the 
full set of CMB data, particularly BOOMERanG, MAXIMA-1, and COBE-DMR.  
\cite{padmanabhans00} analyze the combined BOOMERanG and MAXIMA-1 data 
using a relative likelihood analysis of their $\chi^2$ values; since 
their best-fit model is close to $\chi^2/d.o.f.=1$ this is nearly 
the correct frequentist approach.  The problem is that they ignore 
the significant calibration uncertainties of BOOMERanG and MAXIMA-1 so 
this analysis is seriously flawed and will eliminate a large set of 
viable models.  

An acceptable frequentist analysis of BOOMERanG, MAXIMA-1, and 
COBE-DMR data has been performed by \cite{griffithssz00} in the course 
of adding a bump to the primordial power spectrum in the $\Lambda$CDM 
model.  The standard $\Lambda$CDM model, i.e. amplitude of bump equals 
zero, is ruled out at the 68\% confidence level but not at 95\% confidence.   
This means that $\Lambda$CDM is a reasonably good fit to the 
current set of high-quality CMB data, and it seems to eliminate the 
motivation for considering a primordial bump.  
These authors do not analyze models 
with higher baryon fractions but most likely would find an even better 
fit.  One must then consider whether a model can be ruled out not for 
being a bad fit but simply because another model is a better fit.  In 
general, this is a dangerous approach although it is implicit in 
the Bayesian formalism.  If there is evidence to suggest that the 
observational errors have been overestimated\footnote{How often 
does this happen in observational astrophysics?} then the relative 
likelihood approach may be justified, but otherwise it is premature 
to discard models for which the data is a quite reasonable result.

\section*{Systematic Uncertainties in Baryon Density from Deuterium}

An alternative manner in which the careful usage of statistics may 
resolve the apparent conflict between CMB and BBN values of the baryon 
density is via a fuller accounting of systematic uncertainties 
in the usage of the observed deuterium-to-hydrogen abundance ratio to 
infer the baryon density.  This issue is explored by \cite{burlesnt00}, 
who find that while the best-fit CMB baryon density of $\Omega_b h^2=0.03$ 
``cannot be accomodated,'' a very conservative consideration of 
systematic errors would allow $0.016 \leq \Omega_b h^2 \leq 0.025$.  
Although it is unlikely that the systematic errors in converting the 
observed deuterium-to-hydrogen abundance ratio into a baryon density are 
nearly this large, this does allow for the possibility that improved 
deuterium observations could reduce the claimed $3 \sigma$ conflict 
between the CMB- and BBN-preferred baryon densities to $1 \sigma$.  

Although the above range is quite conservative, the most recent 
high-redshift quasar absorption system in which \cite{omearaetal01} measured 
the deuterium-to-hydrogen abundance ratio, HS0105+1619, would by itself give a 
result of $\Omega_b h^2 = 0.023$.  There are reasons to believe that this 
is the best measurement of deuterium yet performed; it has the 
highest hydrogen column density and therefore deuterium was seen in 
several Lyman-series transitions with a reduced chance of contamination 
from the Lyman alpha forest.  Such contamination would increase the perceived 
abundance of deuterium, leading to an underestimate of the true baryon 
density.  Indeed the HS0105+1619 deuterium-to-hydrogen abundance ratio is 
higher than the two previous detections of \cite{burlest98} by an amount  
greater than the observational errors.  \cite{omearaetal01} are forced 
to add an empirical uncertainty to these points 
in order to account for their scatter.  Unfortunately, it is also possible 
that mild levels of deuterium destruction due to star formation in the 
higher metallicity system HS0105+1619 have caused a systematic error 
in this system instead of the previous ones.  Although deuterium 
destruction is not expected to be significant at the 
significantly sub-solar metallicity of this system, 
it is unclear where the 
true deuterium-to-hydrogen abundance ratio lies amongst the range of 
observed values.  

\section*{Conclusion}

There are thus a number of ways in which a careful usage of statistics 
seems to eliminate the claimed $3 \sigma$ conflict between the CMB- 
and BBN-preferred values of the baryon density.  The first is that the 
Bayesian analyses used are actually producing consistent results; 
the proper conclusion to be drawn is that whether or not to include 
prior information from BBN is an important choice.  Utilizing relative 
likelihood information and prior probability functions makes these analyses 
poorly suited to answering the question of whether a conflict 
exists between the CMB and BBN values for $\Omega_b h^2$.  When a 
better-suited frequentist analysis is used, we find that the standard 
$\Lambda$CDM model with a BBN plus deuterium preferred value of 
$\Omega_b h^2 = 0.021$ is in reasonably 
good agreement with recent CMB observations.  
While a model with higher baryon fraction may be an even better fit this 
could simply be caused by having fit several free parameters; we need 
more precise observations to make a clear discrimination between these 
models.  Additionally it is possible that systematic errors in measuring 
the deuterium-to-hydrogen abundance ratio are responsible for underestimating 
the BBN value of the baryon density.  Given any one of these reasons, there 
is no longer a 
conflict between CMB anisotropy results and the value of $\Omega_b h^2$
preferred by observations of deuterium.  Given all of them, it is clearly 
unnecessary to introduce additional physics to the early universe.





 
\end{document}